# Wide-field 3D nanoscopy on chip through large and tunable spatial-frequency-shift effect

Xiaowei Liu,[1,2,†] Chao Meng,[1,†] Xuechu Xu,[1] Mingwei Tang,[1] Chenlei Pang,[1] Yaoguang Ma,[1] Yaocheng Shi,[1] Qing Yang,[1*] Xu Liu,[1*] and Clemens F. Kaminski[3]

[1]*State Key Laboratory of Modern Optical Instrumentation, College of Optical Science and Engineering, Zhejiang University, Hangzhou 310027, China*

[2]*Guangdong Provincial Key Laboratory of Optical Fiber Sensing and Communications, Institute of Photonics Technology, Jinan University, Guangzhou 510632, China*

[3]*Department of Chemical Engineering and Biotechnology, University of Cambridge, Cambridge, CB30AS, United Kingdom*

[†]*Co-first author: Contributions from X.L. and C.M. are 7:3.*

*Corresponding authors: qingyang@zju.edu.cn, liuxu@zju.edu.cn*

**Abstract**
Linear super-resolution microscopy via synthesis aperture approach permits fast acquisition because of its wide-field implementations, however, it has been limited in resolution because a missing spatial-frequency band occurs when trying to use a shift magnitude surpassing the cutoff frequency of the detection system beyond a factor of two, which causes ghosting to appear. Here, we propose a method of chip-based 3D nanoscopy through large and tunable spatial-frequency-shift effect, capable of covering full extent of the spatial-frequency component within a wide passband. The missing of spatial-frequency can be effectively solved by developing a spatial-frequency-shift actively tuning approach through wave vector manipulation and operation of optical modes propagating along multiple azimuthal directions on a waveguide chip to interfere. In addition, the method includes a chip-based sectioning capability, which is enabled by saturated absorption of fluorophores. By introducing ultra-large propagation effective refractive index, nanoscale resolution is possible, without sacrificing the temporal resolution and the field-of-view. Imaging on GaP waveguide material demonstrates a lateral resolution of $\lambda/10$, which is 5.4 folds above Abbe diffraction limit, and an axial resolution of $\lambda/19$ using 0.9 NA detection objective. Simulation with an assumed propagation effective refractive index of 10 demonstrates a lateral resolution of $\lambda/22$, in which the "huge gap" between the directly shifted and the zero-order components is completely filled to ensure the deep-subwavelength resolvability. It means that, a fast wide-field 3D deep-subdiffraction visualization could be realized using a standard microscope by adding a mass-producible and cost-effective spatial-frequency-shift illumination chip.

Super-resolution microscopy with both fast acquisition and deep-subdiffraction resolution is desirable for practical applications such as visualizing dynamic processes at subcellular resolution in biomedical research [1-3]. A majority of developed super-resolution techniques rely on nonlinear response to achieve nanoscale resolutions [4-13], while their low throughput could limit their practical applications that require high temporal resolution. To achieve fast imaging with a large field-of-view (FOV), linear super-resolution microscopy via synthesis aperture approach is an alternative way. Due to its wide-field implementation, the space-bandwidth-product of the detection system is fully utilized in every frame, which enables the high acquisition speed [2, 14-18].

However, the achievable resolution improvement is limited in the current linear super-resolution microscopy. To the best of our knowledge, its highest reported promotion over diffraction-limited microscopy extends to a 3-fold increase by introducing a near-field illumination module with large illumination wave vectors to shift high spatial-frequencies into the passband of detection system, for example in the surface plasmon polariton-based structured illumination microscopy (SIM) [19-21] and our previously reported nanowire ring illumination microscopy [14, 15]. This weak super-resolution performance arises from that, when trying to use a shift magnitude surpassing the cutoff frequency of the detection system by two folds, a missing spatial-frequency band would occur between the shifted and zero-order components, leading to failed reconstruction and ghosting. Consequently, a large spatial-frequency-shift (SFS) scheme capable of full-coverage detection in the spatial-frequency domain is desirable to achieve a high-throughput, distortion-free, and deep-subwavelength imaging.

An additional feature providing fast, subdiffraction axial sectioning is crucial to locate subwavelength details in three dimensions. Classical subdiffraction sectioning is based on the modulation of evanescent field penetration depth through a precise variation of the incidence angle, which, enables an axial resolution of 40~100 nm, but is complex to implement [22, 23]. A recently reported sectioning method achieved an axial resolution of 20 nm via photobleaching; however, this comes at a severe loss in the acquisition speed because typically $10^5$~$10^6$ absorption/emission cycles occur before a molecule is photobleached [24]. Furthermore, it is difficult to vary the penetration depth in other common evanescent field configurations, such as in chip-based evanescent illumination [14, 15, 25] and localized evanescent illumination [19-21] methods, which results in poor compatibility with existing platforms.

Here, we illustrate a chip-based wide-field nanoscopy method that enables deep-subdiffraction resolution both laterally and axially. To overcome the resolution limitation in the current linear super-resolution microscopy, the broad tunability of shift magnitude in spatial-frequency domain is essential. Interference has been frequently used in microscopy, such as in SIM and interferometric microscopy [26], to produce illumination patterns or to record images. However, little attention has been paid to the broad tunability of the spatial-frequency of the illumination pattern through interference. We propose an SFS tunable nanoscopy (STUN) with large and tunable SFS effect to enable distortion-free, deep-subdiffraction imaging. The spatial-frequency of the illumination pattern can be tuned flexibly through wave vector manipulation and operation in interference using a practical approach of changing the azimuthal incidence direction of the input modes in an SFS waveguide chip. Accordingly, the shift vector can be broadly tuned to achieve full-coverage wide-passband detection in the spatial-frequency domain. It is noteworthy that, this tuning process does not change the penetration depth of the illumination field, and is therefore safe from artefacts when combining multiple frames. Simulation with an assumed propagation effective refractive index of 10 demonstrates a lateral resolution of $\lambda/22$, in which the "huge gap" between the directly shifted and the zero-order components is completely filled in spatial-frequency domain to ensure the deep-subwavelength resolvability.

A novel axial sectioning mechanism is proposed via saturated absorption, which is universal and straightforward to implement. In contrast to the previous sectioning methods in which the number of excitation/emission cycles depends linearly on illumination intensity, nonlinear saturated absorption enables high imaging speed, without negatively affecting the penetration depth. The sectioned layer depth is determined by the incident power, and the sectional image of each layer can be extracted using short pulsed excitations, promising fast acquisition and low photo-toxicity.

Moreover, the methodologies can be implemented through a waveguide-on-chip illumination module, to achieve a chip-based wide-field 3D nanoscopy. As the illumination and detection can be decoupled, it is convenient to enable 3D nanoscopy on a standard microscope by adding a mass-producible illumination chip [14, 15, 25, 27-31]. The manipulation of the propagation wave vector and arbitrarily large FOV can be realized through photonic chip design.

In theory, the resolutions can be significantly improved by introducing ultra-large propagation effective refractive index. Considering the available photonic materials in the visible spectrum, we select GaP waveguide chip in simulations and report a lateral resolution of $\lambda/10$ and an axial resolution of $\lambda/19$ using 0.9-numerical-aperture (NA) detection objective. The lateral resolution is improved 5.4 folds from the Abbe diffraction limit, exceeding the 2~3-fold limited promotion in the previous linear super-resolution microscopy techniques [2, 14-17, 19-21].

Fig. 1 (a) shows the schematic setup. Light is passed from rectangular waveguides into a polygonal waveguide, and interference between optical modes from the different rectangular waveguides creates a structured evanescent illumination pattern for samples atop the waveguide surface. The rectangular waveguides are tens of micrometers in width, so the input modes barely diverge in the polygonal region. The intensity distribution of the interference pattern $I_{illu.}$ can be illustrated by

$$I_{illu.}(\vec{r}) = \left[ A_1(\vec{r})^2 + A_2(\vec{r})^2 \right] + 2A_1(\vec{r})A_2(\vec{r})\cos\left[ (\vec{k}_1 - \vec{k}_2) \cdot \vec{r} + \Delta\varphi \right] \quad (1)$$

where $A_1$ and $A_2$ denote the amplitudes of the two input modes, $\vec{k}_1$ and $\vec{k}_2$ denote their propagation wave vectors, and $\Delta\varphi$ denotes their initial phase difference. $|\vec{k}_1| = |\vec{k}_2| = 2\pi n_{eff.}/\lambda_{illu.}$, where $n_{eff.}$ denotes the effective refractive index, and $\lambda_{illu.}$ denotes the illumination wavelength. The second term of $I_{illu.}$ represents a modulated illumination with a resulting wave vector equal to the superposition of the two input propagation wave vectors. Consequently, the illumination wave vector can be broadly tuned, depending on the intersection angle of the two input modes [Fig. 1(b)].

Overlap between various illumination patterns spans a circular region at the center of the polygonal waveguide, creating an FOV with a diameter approximately equal in width to the input waveguides [red dashed circle in Fig. 1(a)], independent of the NA of a far-field illumination lens. To avoid nonuniform illumination induced by multi-mode propagation and interference, mode filters can be installed upstream for pure fundamental mode operation [32].

Figs. 1(c-d) show the examples of the tunability of the illumination period and direction through wave vector manipulation by selecting different input ports on. The SFS vector can be actively tuned accordingly to collect all the necessary spatial-frequency information. As a result, wide-passband, full-coverage detection in the spatial-frequency domain can be realized by designing the numbers and directions of input ports in the chip-based STUN.

Chip-based STUN is different from classical SIM, although both of them utilize sinusoidal stripe illumination patterns. (*1*) In SIM, the highest SFS magnitude is limited by the NA of the objective. In STUN, the highest SFS magnitude relies on the effective refractive index of the propagating mode in the waveguide, which can be much larger than the NA of the objective. (*2*) In SIM, only the orientation of the illumination pattern is scanned. In STUN, both the orientation and the spatial-frequency of the illumination pattern are scanned to completely cover a much wider passband in the spatial-frequency domain.

Chip-based STUN is also different from solid immersion lens (SIL) imaging, although their resolutions are both closely related to the refractive index of the material. In SIL, there is a tradeoff between resolution and FOV. In STUN, a large FOV and high resolution can be achieved simultaneously.

Fig. 2 shows a comparison of the detectable spatial-frequency ranges in classical SIM, SFS microscopy with a large and fixed shift magnitude, and chip-based STUN with a tunable SFS magnitude. In classical SIM, the SFS magnitude is limited to $2NA/\lambda$, resulting in a cutoff frequency of $4NA/\lambda$. In large SFS microscopy without SFS-tunability, a missing band exists in the spatial-frequency domain between the zero-order component and the shifted component, resulting in failed reconstruction and ghosting. In STUN, the spatial-frequency component within $2(n_{eff.} + NA)/\lambda$ can be completely collected. Considering the illumination wavelength $\lambda_{illu.}$ is shorter than the emission wavelength $\lambda_{em.}$, the resolution improvement can be calculated by

$$(n_{eff.} \cdot \lambda_{em.}/\lambda_{illu.} + NA)/NA \quad (2)$$

For each determined SFS vector, three images with different phases are recorded to solve the zero-order, and the $\pm$ shifted components. Aiming to make the phase modulation more flexible in the future, we randomly set the phases in the simulation, and resorted to a phase estimation reconstructing algorithm to recover the spatial-frequency information [33]. A practical modulation of the phase can be performed outside the chip or by installing a waveguide phase modulator on the chip [34].

We perform simulations to illustrate the lateral resolvability and the FOV. GaP is selected as the waveguide material due to its high refractive index and low loss in the visible spectrum. Its refractive index is 3.74 at a 470 nm vacuum wavelength [35]. The side-length and thickness of the decagonal waveguide are 100 μm and 500 nm, respectively. The effective refractive index of the TM fundamental mode is 3.7. The object is formed by randomly positioning 25-nm-sized fluorophores with a 500 nm emission vacuum wavelength. The NA is set at 0.9, such that a dry objective can be used in this imaging scheme.

Under the illumination scheme defined by the above parameters, the resolution could be improved by 5.4 folds beyond the Abbe diffraction limit, exceeding the typical 2~3-fold limited promotion in the previous linear super-resolution microscopy [2, 14-17, 19-21]. Fig. 3(c) shows the reconstructed STUN image, in which the resolution is dramatically improved compared with that of the wide-field image (Fig. 3(a)) and the classical SIM image (Fig. 3(b)). Two fluorophores with a center-to-center distance of 50 nm could be separated, corresponding to a $\lambda_{em.}/10$ resolution.

The tunability of the SFS magnitude is essential for image reconstruction in large SFS nanoscopy. Otherwise, the missing band in the frequency domain makes it impossible using the classical correlation-based algorithms to estimate the modulation phases of illumination, which, nevertheless, are necessary to solve and shift back the high spatial-frequency components. Moreover, the missing band generates artefacts and distortions. Fig. 3 (e2) shows a simulated image of two 50-nm-spacing fluorophores with large and fixed SFS magnitude. Severe artefacts exist. Besides, the imaged peak-to-peak distance is 60 nm, distorted from the reality.

Chip-based STUN is advantageous for achieving a large FOV, because a low-NA detection objective can be used. However, nonuniform illumination, a common issue in chip-based microscopy due to the mode distribution properties [25, 32], degrades the image quality and causes problems in image processing. To achieve uniform subdiffraction resolution and brightness in a large FOV, we developed a calibration algorithm to reconstruct the STUN image in the entire illuminated area. By calibrating the zero-order component using background intensity distributions and the ± shifted components using modulation depth distributions, spatial-frequency information shown in Fig. 2(c) is sequentially recovered with the SFS magnitude ranging from the lowest to the highest, to reconstruct a deep-subwavelength image with a large FOV and suppressed artefacts. Fig. 3(f) shows several 50-nm-spacing fluorophore pairs placed differently along a diameter in the FOV, which can all be correctly reconstructed.

To further achieve chip-based 3D subdiffraction imaging, we propose a novel axial sectioning method, based on saturated absorption, which is more compatible than previous methods that vary the incidence angle. With this method, optical sectioning can be realized in most types of evanescent illumination, and we demonstrate in principle that chip-based deep-subwavelength sectioning is possible. We illustrate the concept using the setup in Fig. 4(a) (consistent with the model in STUN).

The intensity of the evanescent wave decays exponentially with the distance from the interface. It's possible that the fluorophores are saturated near the film surface but are unsaturated further away [solid blue curve in Fig. 4(b)]. With a larger input intensity, the emission of the saturated fluorophores barely increases [dashed blue curve in Fig. 4(b)], and can be removed by subtraction between the two images [Fig. 4(c)]. As saturation occurs within several nanoseconds, the acquisition time can be short enough for dynamic studies.

Use of excitation with a pulse width shorter than the lifetime of the fluorophore is helpful to avoid photobleaching because of the short duration of electrons staying in the excited state. Considering the condition of single-photon absorption [36], the emission intensity $I_{em.}$ of a fluorophore at an excitation intensity $I_{illu.}$ can be expressed by

$$I_{em.} \propto 1 - e^{-(I_{illu.}\sigma + k_f)t_p} \tag{3}$$

where σ is the absorption cross section, $k_f$ is the fluorescent decay rate, and $t_p$ is the pulse duration. When the excitation intensity increases by $\Delta I_{illu.}$, the increment of $I_{em.}$ can be calculated by

$$\Delta I_{em.} = I_{em.}' \cdot \Delta I_{illu.} \propto e^{-(I_{illu.}\sigma + k_f)t_p} \cdot \Delta I_{illu.} \tag{4}$$

where $I_{em.}'$ is the derivative of $I_{em.}$ to $I_{illu.}$. Under the evanescent wave illumination, $I_{illu.}(z) = I_s \cdot \exp\left(-2\sqrt{n_{eff.}^2 - 1} \cdot \frac{2\pi}{\lambda_{illu.}} \cdot z\right)$, where $I_s$ is the illumination intensity at the surface. Then Equation (4) becomes

$$\Delta I_{em.}(z) \propto \Delta I_s \cdot \exp\left[-I_s \sigma t_p \exp\left(\frac{-4\pi\sqrt{n_{eff.}^2 - 1} \cdot z}{\lambda_{illu.}}\right)\right] \cdot \exp\left(\frac{-4\pi\sqrt{n_{eff.}^2 - 1} \cdot z}{\lambda_{illu.}}\right) \tag{5}$$

The relationship between $\Delta I_{em.}$ and the axial position $z$ follows the curve shown in Fig. 4(c). The peak position, i.e., sectioning depth is located at

$$z = \frac{\ln(N_s) \cdot \lambda_{illu.}}{4\pi\sqrt{n_{eff.}^2 - 1}} \tag{6}$$

where $N_s = I_s \sigma t_p$, which refers to the number of photons that hit one fluorophore at the waveguide surface in a single pulse. The axial imaging depth can be scanned by varying the incident power ($N_s$ or $I_s$), and the axial scanning step size is determined by the power ratios. For each incident power, the sectional image of the corresponding layer can be extracted by subtracting two images under two pulsed excitations with a small change in power. The full-width-at-half-maximum (FWHM) determined axial resolution can be calculated by Equation (7).

$$R_z = \frac{0.2 \lambda_{illu.}}{\sqrt{n_{eff.}^2 - 1}} \tag{7}$$

The axial resolution is determined by the illumination wavelength and the effective refractive index, regardless of the NA of the objective. The shorter excitation wavelength and larger effective refractive index lead to a finer axial resolution.

In the 3D simulation, the object is modeled by positioning 20-nm-sized fluorophores at different depths [Fig. 5(a)]. The illumination and emission vacuum wavelengths are set at 470 nm and 500 nm respectively. Figs. 5(b1-b2) show simulated images with $N_s = 300$ and $N_s = 330$, respectively. Their subtraction sections an ultra-thin layer with a 60 nm distance from the surface [Fig. 5 (b3)]. Another sectioned layer at an 140 nm distance from the surface is achieved at $N_s \sim 6 \times 10^5$ [Fig. 5 (c1-c3)]. Fig. 5(d) shows a 3D reconstruction using the sectional images. FWHM of the axial imaging profile of a fluorophore is 26 nm, corresponding to a theoretical axial resolution of $\lambda_{em.}/19$.

To further improve the axial resolution with a given illumination waveguide chip, multi-photon absorption saturation effect can be introduced. Using absorption of n photons, the axial resolution can be improved by n folds compared to the single-photon absorption condition. In the former case, more axial pixels can be extracted within the penetration depth of the evanescent illumination, and the corresponding power difference is smaller amongst the neighboring pixels, which can suppress the photo-damage to some extent.

In this Letter, we introduced a chip-based STUN method with an optical sectioning capability, which enables deep-subdiffraction resolutions and high imaging speed in both lateral and axial dimensions with a large FOV. The sectioned image can then be segmented using the lateral super-resolution mask for chip-based 3D volumetric nanoscopy [23]. The position of each detected circular sub-range in the spatial-frequency domain can be actively selected by choosing the input ports, and redundant overlap can be suppressed to further increase the imaging speed. Metamaterials with unusual refractive indices could conceivably be introduced to achieve ultra-high SFS values [37] for molecular-scale visualization. A main concern regarding the method is that, although resolution can be unlimitedly improved by increasing the propagation wave vector of the evanescent wave illumination, the resultant shallower field penetration depth would limit the capability for volumetric imaging. Sectioning based on saturated absorption, on the other hand, permits fast acquisition because of the short time required to reach saturation; however, the high excitation intensity brings about the risk of photo-damage to delicate samples. The negative effect might be mitigated through multi-photon saturated absorption. All in all, this design offers a new method of mass-producible, easily-integrated, and high-throughput 3D nanoscope, which could benefit dynamic studies, biomedicine, material science, among others.

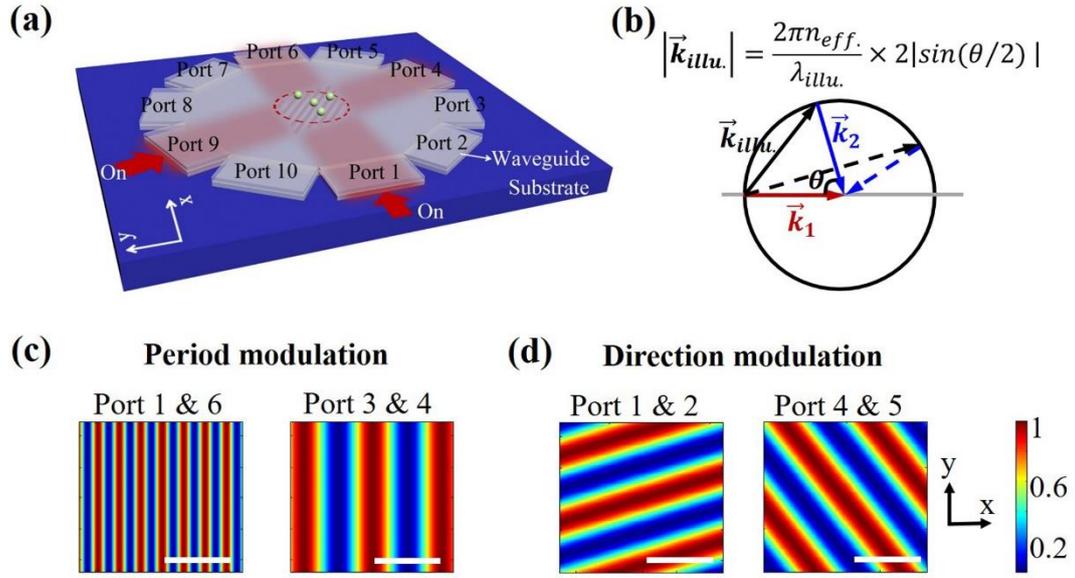

FIG. 1. Mechanism of chip-based STUN. (a) Schematic setup with a decagonal waveguide as an example. The FOV is circled by a red dashed line. (b) Operation of wave vectors. $\vec{k}_1$ and $\vec{k}_2$ denote the input wave vectors, and $\vec{k}_{illu.}$ denotes the illumination wave vector. Solid and dashed arrows represent two different superposition conditions. (c-d) Examples of the interference patterns with different input ports on. Scalebar: 400 nm; $\lambda_{illu.} = 470$ nm; $n_{eff.} = 3.7$.

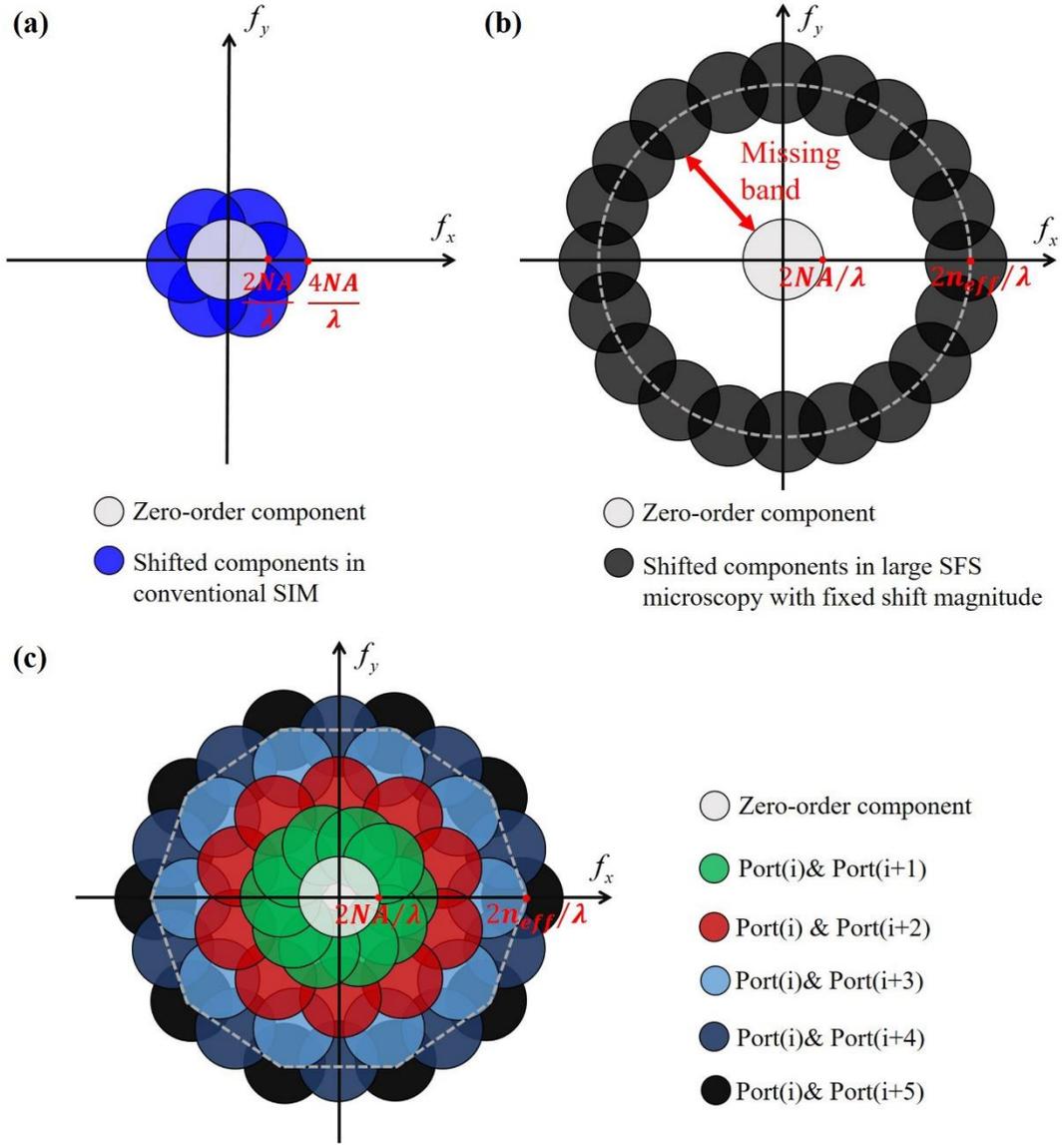

FIG. 2. Detectable spatial-frequency range in classical SIM (a), in SFS microscopy with a large and fixed SFS magnitude (b), and in chip-based STUN (c). Circles of different colors in (c) represent the detectable components with different input ports on. A decagonal waveguide is taken as an example. i = 1, 2, 3, 4, 5; $n_{eff.}$ = 3.7; NA = 0.9.

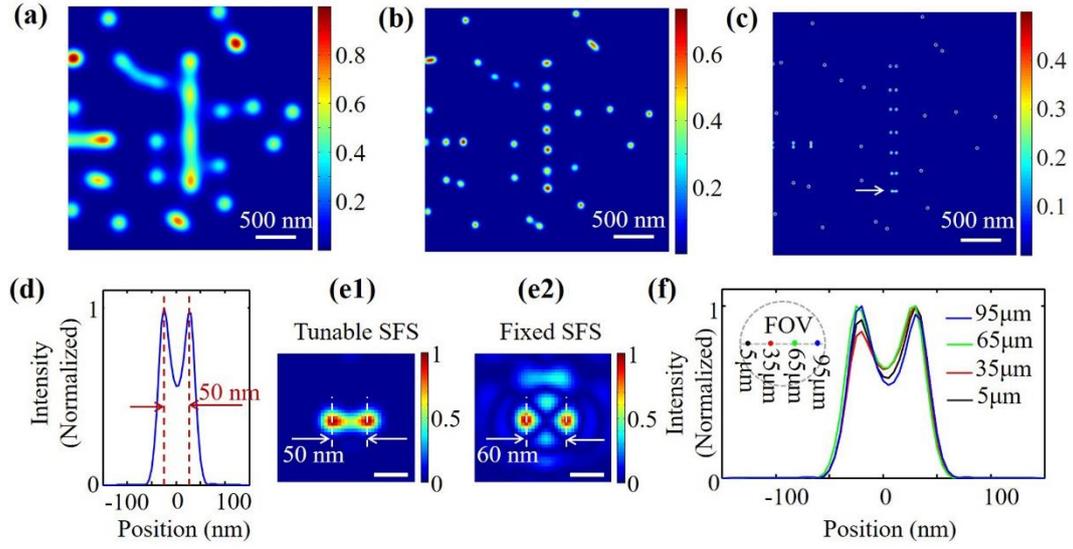

FIG. 3. Resolution and FOV in chip-based STUN. Simulated wide-field image (a), classical SIM image (b), and STUN image (c). (d) Imaging intensity profile of two 50-nm-spacing fluorophores shown through the white arrow in (c). (e) STUN image (e1) and simulated image with large and fixed SFS magnitude (e2). Object: two 50-nm-spacing fluorophores. Scale bar: 50 nm. (f) Image intensity profiles of 50-nm-spacing fluorophore pairs at different positions in the 100-μm-diameter FOV in STUN, as shown in the inset. NA = 0.9; $n_{eff.}$ = 3.7; $\lambda_{illu.}$ = 470 nm; $\lambda_{em.}$ = 500 nm.

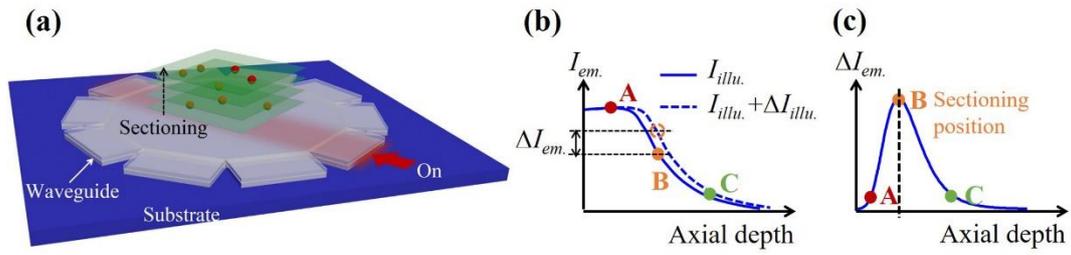

FIG. 4. Sectioning mechanism based on saturated absorption. (a) Schematic of the chip-based sectioning setup. A single port is used for input. (b-c) Emission intensities at different axial depths, for two input intensities $I_{illu.}$ and $I_{illu.} + \Delta I_{illu.}$ (b), and their subtraction (c).

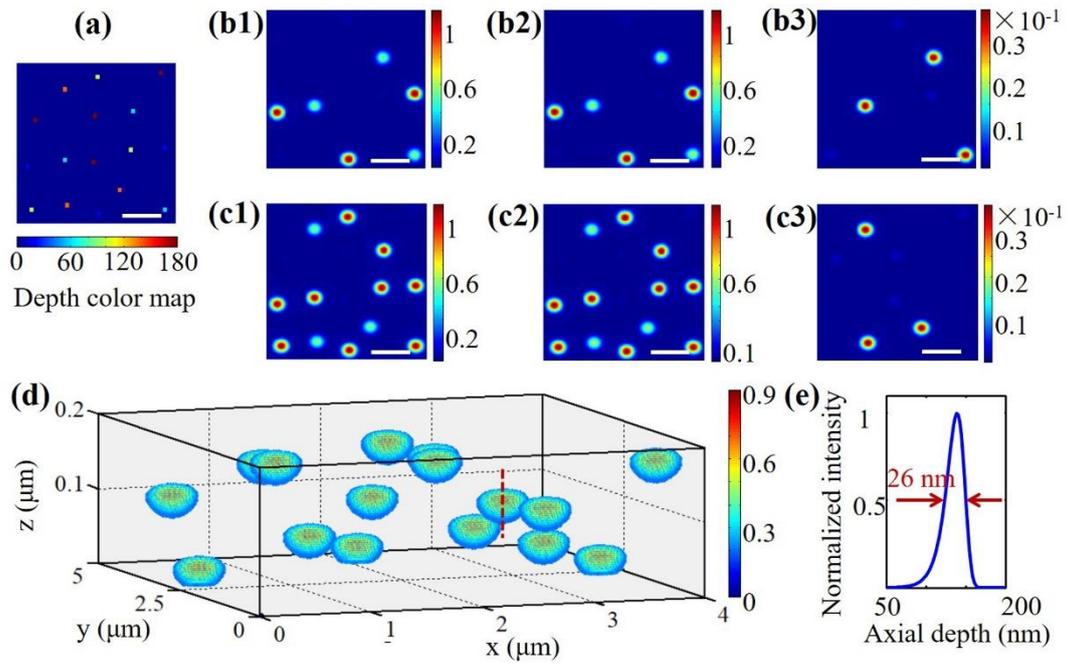

FIG. 5. Simulated sectional images based on saturated absorption. (a) The 3D object. (b-c) Images captured at different illumination intensities and the corresponding subtracted images (b3, c3). Scalebar: 1 μm. (d) Reconstructed 3D image. (e) Image intensity profile plotted along the red dashed line in (d). NA = 0.9; $n_{eff.}$ = 3.7; $\lambda_{illu.}$ = 470 nm.

**Acknowledgments**

We thank Dr. Shoubao Han and Dr. Jingye Chen for the discussions on the waveguide design. This work is supported by National Natural Science Foundation of China (No. 61735017, 61822510, 51672245), Zhejiang Provincial Natural Science of China (No. R17F050003), National Key Basic Research Program of China (No. 2015CB352003), the Fundamental Research Funds for the Central Universities, the Program for Zhejiang Leading Team of S&T Innovation, the Cao Guangbiao Advanced Technology Program, and First-class Universities and Academic Programs.